\documentstyle[preprint,aps]{revtex}

\begin{document}
\title{Slow and fast dynamics in coupled systems: \\
A time series analysis view}

\author{G. Boffetta$^{1}$, A. Crisanti$^{2}$, F. Paparella$^{3}$,
A. Provenzale$^{3}$ and A. Vulpiani$^{2}$}

\address{$^{1}$Dipartimento di Fisica Generale,
Universit\`a di Torino, via Pietro Giuria 1, 10125 Torino, Italy \\
Istituto Nazionale di Fisica della Materia, Unit\`a di Torino}

\address{$^{2}$Dipartimento di Fisica, Universit\`a di Roma ``La Sapienza'',
         p.le Aldo Moro 2, 00185 Roma, Italy \\
Istituto Nazionale di Fisica della Materia, Unit\`a di Roma}

\address{$^{3}$Istituto di Cosmogeofisica del CNR,
         corso Fiume 4, 10133 Torino, Italy}

\date{\today}

\maketitle

\begin{abstract}
We study the dynamics of systems with different time
scales, when access only to the slow variables is allowed. We use
the concept of Finite Size Lyapunov Exponent (FSLE) and consider 
both the case when the equations of motion for the slow components 
are known, and the situation when a scalar time series of
one of the slow variables has been measured. A discussion on the
effects of parameterizing the fast dynamics is given.
We show that, although the computation of the largest
Lyapunov exponent can be practically infeasible in complex
dynamical systems, the computation of the FSLE allows  to extract
information on the characteristic time and on the predictability of the
large-scale, slow-time dynamics even with moderate statistics and
unresolved small scales.
\end{abstract}

\section{Introduction}
\label{sec:1}
In the last two decades, the problem of extracting 
information from a measured time series has been studied 
extensively, see e.g. 
\cite{TO,OSY,Aba,Kant,ER85,Pack,Tak,GP,GPe,Wolf,Smith,ER92}.
Several attempts have been devoted to the issue of 
distinguishing between deterministic and stochastic behavior, 
where ``deterministic" has to be interpreted as ``dominated by a
small number of excited modes" and ``stochastic" as ``dominated by a 
large number of excited degrees of freedom".
Once assessed the presence of low-dimensional chaotic dynamics, various 
methods have been devised for determining the statistical properties 
of the attractor and to build appropriate models for either predicting 
or describing the system evolution. 

Most methods for determining dynamical properties from measured signals 
are based on a procedure of phase-space reconstruction. Following the 
work of  Packard et al \cite{Pack} and Takens \cite{Tak}, the so-called 
time-embedding techniques have been developed to address this problem. 
Their use (e.g. via delay coordinates) allows, at least in principle,
the determination of the dimensions \cite{GP}, the Kolmogorov-Sinai 
entropy \cite{GPe} and the Lyapunov exponents \cite{Wolf,ER85} of the 
system by the analysis of a time series of just one scalar variable. 

Unfortunately, this approach may have severe limitations in many 
practical situations. For example, the length of the time series
is a crucial point in order to obtain reliable estimates of the
phase-space properties of the system \cite{Smith,ER92}.
Further, there are simple stochastic processes that mimic
a ``false positive" answer to the search for low-dimensional
chaotic dynamics, providing a finite value of the dimension
under time-embedding in most practical cases \cite{OP,Vio,Pro}.
Analogously, simple systems characterized by on/off intermittency 
require additional care in the procedure of phase-space reconstruction 
and analysis \cite{PST,Harden}.

Another problem is encountered in systems with many different time 
scales.
In this case, it has been shown \cite{Bof,ABCPV96} that the Lyapunov 
exponents may have a rather marginal role. The growth of a non-infinitesimal 
perturbation is indeed ruled by a non-linear mechanism which depends on 
the details of the system. For this reason, despite the positiveness of 
the largest Lyapunov exponent, it is possible to have a long 
predictability time for some specific degrees of freedom.
A typical example of this type of behavior is provided by 3D
turbulence, 
that is characterized by the contemporary presence
of a hierarchy of eddy turnover times.
In this case, large-scale motions have a predictability time
that is much larger than the one suggested by the value of the 
largest Lyapunov exponent.
In such a situation, the predictability time for realistic 
perturbations may thus have no relationship with the growth rate of infinitesimal perturbations.

As an attempt to overcome this problem, the concept of maximum
Lyapunov exponent has recently been generalized in \cite{ABCPV96} 
to the case of non-infinitesimal perturbations, introducing the notion 
of the Finite Size Lyapunov Exponent (FSLE).
In this work, we further elaborate on this issue and apply 
this method to the detection of ``large scale'' (``slow'')
dynamical properties of measured systems characterized by the 
contemporary presence of different time scales.
In particular, we are concerned with systems 
that can be separated into a slow part, $S$, described by  the phase-space
variables ${\bbox x}_{s}$, and a fast part $F$, described by the variables 
${\bbox x}_{f}$. The two subsystems are coupled through a term of
typical strength $\epsilon$. In the limit $\epsilon\to 0$,
each of the two subsystems evolves independently with its own (chaotic) dynamics. The Lyapunov
exponents of the slow and fast subsystem are $\lambda_{s} < \lambda_{f}$ respectively.

As for the coupling, we can either have
a situation where the fast subsystem drives the slow one without 
being influenced by the latter, 
see for example\cite{PST,Harden}, 
or a more generic coupling between the 
two parts \cite{Bof,BGPV97}. For the specific application we are concerned 
with here, the form of the coupling is not very important.
Preliminary results on the predictability of a slow system $S$ coupled with a faster system $F$ 
have been discussed in the case of two coupled Lorenz models 
\cite{BGPV97}. In that work, the dynamics of the fast system was 
supposed to be known with arbitrary accuracy, and 
it was found that even if the value of the Lyapunov exponent is
determined by the fast dynamics, the predictability of the slow system 
is dominated by its own characteristic time and it is almost unaffected 
by a small coupling with the fast dynamics. 

Physically, we may think of the fast subsystem as representing 
small scales
that, both in real experiments and numerical simulations, are not
resolved. Consistent with this interpretation, 
here we assume that the dynamics of the 
fast subsystem is poorly known, and investigate the effects of 
parameterizing the fast dynamics
when one has access only to the slow dynamics.
In this framework, we consider two different situations. 
In the first case, the equations of motion of the slow system 
are given. In the second case, we consider the computation of 
the FSLE directly from the a measured time series.
The study of systems with two characteristic time
scales is the first necessary step for the understanding of more
realistic systems with several scales \cite{Lorenz96}.

The remainder of this paper is organized as follows.
In Section 2 we discuss the notion of Finite Size Lyapunov Exponent
introduced in \cite{ABCPV96}. In Section 3 we study the case of 
two coupled systems having different
time scales, when access to the whole phase space
of the slow system is allowed. In Section 4 we consider the same cases,
but when just one scalar time series is supposed to have been
measured. Section 5 gives conclusions and perspectives.

\section{Extension of the Lyapunov exponent to finite perturbations}
\label{sec:2}
Here we recall the basic ingredients and the algorithm for computing the 
Finite Size Lyapunov Exponent (FSLE), referring to \cite{ABCPV96} for further
details.
The definition of FSLE follows from that of error growing time $T_r(\delta)$
for a perturbation of size $\delta$. By definition, $T_r(\delta)$ is the 
time that a perturbation with initial size $\delta$ takes to grow by a factor 
$r$ during the system evolution.
In general, the perturbation with size $\delta$ is supposed to be already 
aligned with the most unstable direction.
The error ratio $r$ should not be taken too large, in order to avoid the 
growth through different scales. In many applications, $r=2$, 
so sometimes the $T_r$ is also called the error doubling time.
The Finite Size Lyapunov Exponent is defined from an ensemble 
average of predictability time according to
\begin{equation}
\lambda(\delta) =  
{1 \over \langle T_r(\delta) \rangle} \ln r = 
\left\langle {1 \over T_r(\delta)} \right\rangle_{t} \ln r
\label{eq:2.1}
\end{equation}
where $\langle ... \rangle_{t}$ denotes the natural measure along the
trajectory and $\langle ... \rangle$ is the average over many realizations. 
The second equality comes from the definition of the time average along a
trajectory for a generic quantity $A$:
\begin{equation}
  \langle A \rangle_t = {1\over T} \int_0^T A(t)\, dt
   = {{\sum_i A_i\, \tau_i}\over{\sum_i \tau_i}}
   = {{\langle A\,\tau \rangle}\over{\langle \tau\rangle}}.
\label{eq:av}
\end{equation}
in the particular case of $A=1/\tau$ \cite{ABCPV96}.

In the limit of infinitesimal perturbations, $\delta \rightarrow 0$, 
this definition reduces to that of the leading Lyapunov exponent 
$\lambda_{\rm max}$.
In practice, $\lambda(\delta)$ displays a plateau at the value 
$\lambda_{\rm max}$ for sufficiently small $\delta$.

In most systems, the smaller scales evolve faster, as in the classic 
example of three dimensional turbulent flows, and dominate 
the error growth for infinitesimal perturbations. 
When the size $\delta$ of the perturbation cannot be considered any 
longer infinitesimal, all the scales whose typical size 
is smaller than $\delta$ experience a diffusive separation and do not
contribute to the exponential divergence in phase space. 
At this stage, the behavior of $\lambda(\delta)$ is governed by the 
nonlinear evolution of the perturbation, and, in general,
$\lambda(\delta) \leq \lambda_{\rm max}$. The decrease of 
$\lambda(\delta)$ does follow a system-dependent law. 
In some cases, $\lambda(\delta)$ can be predicted by dimensional
considerations. For the fully developed turbulence, for example,
dimensional considerations lead to the universal law $\lambda(\delta)
\sim \delta^{-2}$ in the inertial range \cite{ABCPV96}.

Therefore, the behavior of $\lambda$ as a function of $\delta$
contains important informations on the characteristic times governing 
the system, and it is a powerful tool for investigating the behavior of
high-dimensional dynamical systems involving many characteristic 
scales in space and time.

To practically compute the FSLE, one has first
to define a series of thresholds $\delta_{n}=r^{n} \delta_{0}$, and to
measure the time $T_r(\delta_{n})$ that a perturbation with 
size $\delta_{n}$ takes to grow up to $\delta_{n+1}$. The time
$T_r(\delta_{n})$ is obtained by following the evolution of the 
perturbation from its initial 
size $\delta_{\rm min}$ up to the largest threshold $\delta_{\rm max}$.
This can be done, for example, by integrating two trajectories of the 
system that start at an initial distance $\delta_{\rm min}$.
In general, one must take $\delta_{\rm min}\ll\delta_{0}$,
in order to allow the direction of the initial perturbation
to align with the most unstable direction in the phase-space. 
The FSLE, $\lambda(\delta_{n})$, is then computed by averaging the 
predictability times over several realizations, see equation
(\ref{eq:2.1}). 

Note that the  FSLE has conceptual similarities with
the $\epsilon$-entropy \cite{Kol56}. This latter measures 
the bandwidth that is necessary for reproducing the trajectory of
a system within a finite accuracy $\delta$. The $\epsilon$-entropy approach has
already been applied to the analysis of simple systems and experimental
data \cite{GW93}, giving interesting results. The direct calculation of
the $\epsilon$-entropy, however, is much more expensive than that 
of the FSLE. 
This latter, in fact, is not more expensive than that of
the largest Lyapunov exponent $\lambda_{\rm max}$.

\section{FSLE and small scale parameterization}
\label{sec:3}
In this section we study the case 
of a slow system $S$, described by the
variables ${\bbox x}_{s}$, coupled 
with a fast system $F$ described by the variables 
${\bbox x}_{f}$. The equations of motion governing 
the slow variables are supposed to be known, and we study the 
effects of parameterizing the fast dynamics.

To study the evolution of the perturbation,
we consider two trajectories $\bbox{x}=({\bbox x}_{s},{\bbox x}_{f})$
(reference) and $\bbox{x}'$ (perturbed) starting from two nearby locations
in phase space. 
The perturbed trajectory is then made to evolve
according either to the same equations as
the reference one, or to modified equations where the
fast dynamics is replaced by a stochastic process, or
simply neglected.

\subsection{Coupled maps}
The first example is provided by two coupled maps, namely
\begin{equation}
\label{eq:ThreeMap}
 \left\{ \begin{array}{lcl}
  x_s(n+1) &=& (1-\epsilon)\,f_s[x_s(n)] + \epsilon\,g[x_s(n),x_f(n)], 
                \qquad {\rm mod}\,1 \\
  x_f(n+1) &=& (1-\epsilon)\,f_f[x_f(n)] + \epsilon\,g[x_f(n),x_s(n)],
                \qquad {\rm mod}\,1
  \end{array}
  \right.
\end{equation}
where $f_s$ and $f_f$ are maps of the unit interval $[0,1]$ into 
itself. Here we use
\begin{equation}
\label{eq:ThreeMap1}
         f_s(x_s) = \left.e^{\lambda_s}\,x_s\right|_{{\rm mod}\,1};\qquad
	f_f(x_f) = \left.e^{\lambda_f}\,x_f\right|_{{\rm mod}\,1};\qquad
	g(x_s,x_f) = \cos\left(2 \pi (x_s+x_f)\right)
\end{equation}
with $\lambda_s<\lambda_f$. Equations (\ref{eq:ThreeMap}) completely 
define the  dynamics of the system. We assume, however, that we can 
have access only to the slow variable $x_s(n)$. The 
FSLE has thus to be computed only from the time 
evolution of $x_s(n)$. 

Before discussing how this can be achieved, we note that equation 
(\ref{eq:2.1}) is inadequate in the case of maps. Definition 
(\ref{eq:2.1}), in fact, tacitly assumes that we are able to determine
the time when the size of the perturbation 
is {\it exactly equal} to the fixed
threshold $\delta$. In the case of maps, this may not be possible. 
The appropriate definition in this case thus becomes
(see \cite{ABCPV96})
\begin{equation}
\label{eq:app2}
  \lambda(\delta) = \frac{1}{\langle n_r\rangle}\, 
                     \left\langle \ln\left( \frac{\delta (n_r)}{\delta}\right)
                     \right\rangle.
\end{equation}
where $\delta$ is the initial size of the perturbation and
$\delta (n_r)$ is its size at the (discrete) time $n_r$. 
Here $n_r$ is the time at which the size of the perturbation
first gets larger than (or equal to) 
$r\delta$, i.e., $\delta(n_r) \ge r\delta$ 
and $\delta(n_r-1) < r\delta$. The average 
$\langle \cdots\rangle$ is over an ensemble of many realizations,
as in the original definition (\ref{eq:2.1}).

Let us now discuss how to compute $\lambda(\delta)$ from the
knowledge of $x_s(n)$. From a point $(x_s,x_f)$ on the system's 
attractor, we generate a new point representing
the perturbed trajectory ($\delta_{\rm min} \ll 1$)
\begin{equation}
\label{eq:NewPoint}
   x_s' = x_s + \delta_{\rm min}, \qquad
   x_f' = x_f + \delta_{\rm min}
\end{equation}
and iterate the coupled maps for the original trajectory and the perturbed one. 
Note that, in this case, the perturbation has been
applied to both the slow and the fast variables. We then
compute $\lambda(\delta)$ from eq. (\ref{eq:app2}).

Figure 1 shows the value of 
$\lambda(\delta)$ versus $\delta$ for the coupled maps system.
The curve with filled triangles has been obtained by defining
the distance in phase space as $\delta=|x_s(n)-x_s'(n)|$. The curve
denoted by the filled squares has been obtained with a different 
definition of distance, namely $\delta=\{[x_s(n)-x_s'(n)]^2+
[x_s(n-1)-x_s'(n-1)]^2\}^{1/2}$ (reminiscent of the time embedding
procedure, see next section). 
Both curves are obtained by an average over
$10^4$ samples for each value of $\delta$; analogous results
are obtained with more limited statistics.
For small $\delta$, the dynamics of the perturbation is driven by the
fast mode, and $\lambda(\delta)$ tends toward 
$\lambda_{\rm max} \simeq \lambda_f=0.5$. For large values of 
$\delta$, the growth of the perturbation is governed mainly 
by the slow dynamics and $\lambda(\delta)$ approaches $\lambda_s=0.1$.
The transition between the two regimes takes place at 
$\delta\sim\epsilon=2 \times 10^{-3}$; changing the strength of the coupling
modifies only the value of $\delta$ where
the transition takes place. 
It is worth noting that the two definitions of distance used in
obtaining figure 1 give almost coincident results. This indicates that,
at least in this case, the definition of distance which is employed is 
not crucial, at variance with what happens in other cases
(e.g. for on/off intermittent systems \cite{Harden}). 

As discussed above, we have always used 
$\delta_{\rm min} \ll \delta_{0}$ 
in order to allow the direction of the initial perturbation
to align with the most unstable phase-space direction, and hence to 
be able to recover the largest Lyapunov exponent 
in the limit of infinitesimal perturbations.
In figure 2 (curve with filled triangles) we show
what happens when this prescription is 
relaxed. 
In this case, the value of $\lambda(\delta)$ is
underestimated for $\delta\sim\delta_{\rm min}$. This
is due to the fact that
the perturbation is not along the most unstable direction.
The process of alignment of the perturbation along the unstable 
direction may result in a decrease of the distance between the 
two trajectories at initial times (i.e., at small
$\delta$), leading to a lower value of $\lambda(\delta)$. The computation
of $\lambda(\delta)$ without relaxation can indeed be more appropriate
for characterizing short time predictability with large initial error,
but we cannot expect, in this case, to asymptotically 
recover the largest 
Lyapunov exponent.

  From these results, it is apparent that the effects of the 
fastest dynamics are seen only when the size of the perturbation is 
small enough. As a consequence, one has that the predictability for 
finite-size perturbations may be unaffected by the particular 
parameterization of the faster scales. To confirm this inference, 
in figure 3 we show  $\lambda(\delta)$ for a case where in the
evolution of the perturbed trajectory $(x_s',x_f')$ 
the fast variable $x_f'$ is
replaced by a sequence of random numbers 
uniformly distributed in the interval
$[0,1]$ (curve with filled triangles). 
The fact that $x_f'$ is now 
a random variable can be detected only for small
enough $\delta$ (where $\lambda(\delta)$ is ruled by a logarithmic 
law, see \cite{ABCPV96}). The 
characterization of the large scale dynamics, however, 
is unaffected by the incorrect parameterization of the fast dynamics. 

Note, also, an interesting point. The results in figure 1 indicate that
the small-scale dynamics is correctly recovered when a 
one-dimensional time series is used ($x_s(n)$), even though 
the full dynamics is two-dimensional ($x_s(n),x_f(n)$). This event,
possibly surprising at first sight, is due to the fact that the perturbed
trajectory has been obtained by acting on the full phase space of the
system; i.e., the perturbation has been made on both $x_s(n)$ and
$x_f(n)$. When only $x_s(n)$ is perturbed, as in the curve shown in
figure 3, it is not possible to recover the fast small-scale dynamics
without resorting to methods such as the time embedding technique.
In this case, however, other problems appear, as discussed in
the next section.

\subsection{Coupled Lorenz models}
To illustrate the application of the FSLE technique to the case of 
continuous-time dynamical
systems, here we consider a system obtained by coupling
two Lorenz \cite{Lorenz63} models having time scales that differ
by a factor $a$.

The slow subsystem is coupled through the Rayleigh 
number $R$ to the
fast one; for simplicity, the fast subsystem does not feel any
feedback from the slow one. More generic (small) couplings do not 
qualitatively change the results, see \cite{BGPV97}. 
The equations for the whole system are
\begin{equation}
 \left\{
 \begin{array}{lcl}
   {d x_s \over dt} &=& -\sigma x_s + \sigma y_s \\
   {d y_s \over dt} &=& -x_s y_s + (R+\epsilon z_f) x_s - y_s \\
   {d z_s \over dt} &=& x_s y_s - b z_s \\
   {d x_f \over dt} &=& (-\sigma x_f + \sigma y_f) \cdot a \\
   {d y_f \over dt} &=& (-x_f y_f + R x_f - y_f) \cdot a \\
   {d z_f \over dt} &=& (x_f y_f - b z_f) \cdot a
 \end{array}
 \right.
\label{clor}
\end{equation}
where the parameter $\epsilon$ controls the  strength of the coupling
and $a$ the relative time scale.
 
In the first type of simulations, both the reference 
and the perturbed trajectories evolve with the same equations of 
motion (\ref{clor}). 
Again, we assume that only the slow variables are 
accessible; the norm is defined as the Euclidean distance in 
the three dimensional space $(x_s,y_s,z_s)$.

The parameters used in the numerical integrations are $R=45.92$, $\sigma=16$,
$b=4$; the Lyapunov exponent of the slow subsystems
is $\lambda_s \simeq 1.5$. The ratio of the time scales of the two
systems is $a=5$, hence 
$\lambda_f \simeq \lambda_{\rm max} \simeq 7.5$. 
The results for $\epsilon=10^{-4}$ are shown in figure 4.
As in the case of the coupled maps, $\lambda(\delta)$ displays two
plateaus at $\lambda(\delta)\simeq\lambda_f$ and
$\lambda(\delta)\simeq\lambda_s$, corresponding respectively
to the fast and slow dynamics, and a transition region
at $\delta\sim\epsilon$. Again, the small-scale dynamics (associated 
with the fast variables) can be recovered because we have perturbed
the trajectory in the full, six-dimensional, phase space.

To investigate the role of small scale parameterization, also for this
system we have considered a situation where 
the ``true" dynamics of the fast variables 
$x_f',y_f',z_f'$ in the perturbation is replaced by
a stochastic process, i.e.
\begin{equation}
 \left\{
 \begin{array}{lcl}
   {d x_s' \over dt} &=& -\sigma x_s' + \sigma y_s' \\
   {d y_s' \over dt} &=& -x_s' y_s' + (R+\epsilon \eta) x_s' - y_s' \\
   {d z_s' \over dt} &=& x_s' y_s' - b z_s' \\
 \end{array}
 \right.
\label{noiselor}
\end{equation}
where $\eta$ is a Gaussian white noise process with variance equal to 
that of the fast component in the original system (\ref{clor}).
Analogously, we have considered a case where the fast dynamics of
the perturbation is simply neglected; this corresponds to taking
$\epsilon=0$ in the evolution equations for the perturbation
$x_s',y_s',z_s'$. The two corresponding curves of $\lambda(\delta)$
are shown in figure 4 (filled triangles and open diamonds).

The results shown in figure 4 confirm that the
estimate of $\lambda(\delta)$ for the slow variables is practically 
unaffected, for large values of $\delta$, by the 
details of the fast dynamics. A similar result was obtained in
\cite{CR} for the chaotic or stochastic resonance of
a driven nonlinear oscillator. 
In the present situation,
one may even neglect the fast dynamics, and still obtain
a reliable estimate of the slow evolution (and of the
Lyapunov exponent associated with the slow variables). 
In particular, the lack of knowledge of the 
fast dynamics has an effect which is similar 
to that associated with the presence of noise. The 
inadequacy of the parameterization for scales smaller than $\epsilon$
is is reflected, in both cases, in a large value of the FSLE, i.e., in a
poor predictability of the phase-space dynamics on small scales. 
At larger scales and slower times, the FSLE coincides
with the Lyapunov exponent of the (uncoupled) slow subsystem. This
defines the predictive skill of the ``incomplete model'' on those scales.

\section{Computation of the FSLE from measured data}
\label{sec:4}
In the case of measured time series, it is not usually possible 
to have access to the whole set of variables describing the system.
Consistent with these limitations, here we 
suppose that only one time series of a scalar
observable quantity $h_n$, function of the slow phase-space 
variables of the system, is given. Additionally, in most 
experimental situation, the time 
series of $h_n$ is characterized by limited statistics.

The first step is the procedure of phase-space reconstruction.
The time-embedding method \cite{Pack,Tak} allows to reconstruct a pseudo 
phase space with dimension $M$, by using time delay coordinates of the 
observed variable.
A vector in this phase space is then defined as: 
\begin{equation}
\bbox{X}_n=\left(h_n,h_{n-\tau},\ldots,h_{n-(M-1)\tau}\right)
\end{equation}
where $\tau$ is a suitably chosen time delay, see e.g.
\cite{OSY,Aba} for a discussion on the optimal choice of $\tau$.
The method for computing the experimental FSLE is
then a simple modification of the standard algorithm for the Lyapunov
exponent \cite{Wolf} which measures the average separation between 
trajectories in the embedding space. 

For each reconstructed vector $\bbox{X}_n$, its nearest neighbor $\bbox{X}_m$
is determined. If the separation $\delta=|\bbox{X}_n-\bbox{X}_m|$
is smaller that a given threshold
$\delta_{\rm min}$, the trajectories starting from $\bbox{X}_n$ and
$\bbox{X}_m$ are used to compute the FSLE,
according to the algorithm discussed above.
As in the case of maps, the trajectory is not continuous in time
and one has to adopt the definition (\ref{eq:app2}).

Also in this case we require $\delta_{\rm min}$ to be considerable smaller
than $\delta_0$, to allow the vector separating the two trajectories to
align with the maximally expanding direction. Clearly, this may severely 
limit the available statistics.
A trivial geometrical argument shows that the 
probability of finding two
points at a given distance in the embedding space becomes 
extremely small for high embedding 
dimensions. We have found this lack 
of statistics to be the most important limitation 
that prevents from taking the limit $\delta
\rightarrow 0$, and thus from estimating the largest Lyapunov 
exponent. On the
other hand, the statistics grows with the perturbation 
threshold $\delta_j$, 
we may thus expect to be able to compute $\lambda(\delta_j)$ for
sufficiently large values of $\delta_j$. 
Another crucial point is related to the fact that
the number of degrees of freedom which participate in the slow
dynamics is (usually much) smaller than the total number of
excited modes. Thus, the embedding dimension which is required for
estimating $\lambda(\delta)$ for moderate values of $\delta$ is 
smaller that than needed for estimating the largest Lyapunov 
exponent.

As a first example, we consider the case where the signal $h_n$ is generated by 
the component $x_{s}(n)$ of the coupled maps model (\ref{eq:ThreeMap}).
The coupling parameter is herein chosen to be 
$\epsilon=0.02$ (larger than that used in the previous section), in 
order to be able to study the small-scale behavior
at $\delta < \epsilon$ 
even with moderate statistics. The results are shown
in figure 5. The three curves show the FSLE as obtained in the
previous section (filled squares), and that from the 
time-embedding method with embedding
dimensions $M=1$ (filled triangles) and $M=2$ (open diamonds)
with time delay $\tau=1$. 
As expected \cite{Pack,Tak}, the
computation of the Lyapunov exponent requires in this case
an embedding dimension
$M=2$ in order to resolve the fast dynamics. In fact,
at variance with the results shown in the previous section,
the one-dimensional
time series is not enough here because we have no control on the
perturbation in the fast variable.
It is interesting, however, that one can obtain 
$\lambda(\delta)$ for large $\delta$ already with an embedding 
dimension which reflects the dimensionality of the slow
system ($M=1$). 

Note that, 
in order to have good statistics, each point if figure \ref{fig:5}
has been obtained by averaging (\ref{eq:app2}) 
over $10^4$ samples. This
requires a time series of about $10^8$ points for $M=1$ and more than 
$10^{10}$ points in the case $M=2$ to resolve the small scales at 
$\delta \sim 10^{-4}$. Thus, although it is in principle
possible to extract information
on the fast dynamics and on the largest Lyapunov exponent 
from a measured time series, the statistics required is so 
prohibitive that this procedure may be infeasible in realistic
situations.
On the other hand, it is possible to extract
information on the large-scale Lyapunov exponent with an embedding
dimension of the order of the number of degree of freedom involved 
in the slow dynamics.

We observe that this result could not be obtained by simply neglecting
the fast component as in a measured time series one has no direct
access to the equations of motion.

As a second example, we now apply 
the same machinery to the case of the
two coupled Lorenz systems described by equation (\ref{clor}).
Figure \ref{fig:6} shows the results obtained by using 
the variable $x_s$ of
(\ref{clor}), with $M=3$, $\tau=0.02$ and a total number of 
$N=500,000$ points in the time series.
The coupling constant between the models is now  $\epsilon=0.05$.
The perturbation threshold is fixed as $\delta_{min}=0.005$
and $\delta_0=0.05$.
The plateau corresponding to the large scales is clearly visible,
while, in spite of the large number of points, the contribution of the
fast system is not resolved. Clearly, this
would require a larger embedding dimension and an
increase of the smallest resolved value of $\delta$, at
the cost of an unrealistic increase of the number of points
in the time series.

\section{Conclusions}
\label{sec:5}

In this work we have discussed how to use the method of the Finite 
Size Lyapunov Exponent (FSLE) for determining the slow dynamics
of systems with many different characteristic times.
In particular, we have considered the case when full access to the
slow dynamics is allowed, and the more realistic case when just
one scalar time series of a slow variable has been measured.
The basic idea is to compute the FSLE, $\lambda(\delta)$, as
a function of $\delta$ in the framework of the embedding technique.
In this case, the behavior of $\lambda(\delta)$ at large values of
$\delta$ gives information on
the Lyapunov exponents associated with the slow dynamics.

By contrast, the behavior of $\lambda(\delta)$  
for small value of  $\delta$ gives information on the fast dynamics.
By considering the limit of $\lambda(\delta)$ for $\delta \to 0$,
it is possible, at least in principle,  to discriminate  between
``stochastic'' systems and ``chaotic but deterministic'' ones.
For an accurate estimate of $\lambda(\delta)$ at small $\delta$'s, 
however, it is necessary
to use a very large number of points in the time series. This fact
makes practically infeasible the calculation of the largest Lyapunov
exponent (associated with the fast dynamics) in most complex
dynamical systems.

Nevertheless, very often the slow, large-scale dynamics is the most
interesting one physically. 
The results obtained here indicate that
the slow dynamics may be satisfactorily detected even 
with a limited number of points and a moderate 
embedding dimension.
In these systems, one can thus obtain a satisfactory prediction 
for the slow, physically interesting scales even when access to 
the (much more unpredictable) 
fast scales is not available. This also indicates that,
at least in the examples considered here 
the parameterization of the fast time scales
seems not to be crucial, as the internal dynamics 
of the slow modes plays the dominant role.

One could wonder, then, how general the results presented in this paper
are. Previous works on more complex theoretical models \cite{BGPV97}
indicate that the crucial point is not the dimensionality of the system
or the details of the couplings, but rather the existence of well
separated, weakly interacting, scales. In the present work we have shown
that the FSLE technique may be successfully applied also in the case of
measured time series.

\section*{Acknowledgments}
This work has been partially supported by the CNR
research project ``Climate Variability and Predictability'' and by the
INFN ``Iniziativa Specifica Meccanica Statistica'' FI11. G.B. thanks the
Istituto di Cosmogeofisica del CNR for hospitality.



\begin{figure}[hbt]
\caption{
The FSLE, $\lambda(\delta)$, as a function of $\delta$ for the
coupled maps (\protect\ref{eq:ThreeMap}, 
\protect\ref{eq:ThreeMap1})
with $\lambda_s=0.1$, $\lambda_f=0.5$ and 
$\epsilon = 2\times 10^{-3}$. 
The two curves refer to the different definitions 
of distance discussed in the text. The parameters 
of the perturbation are
$\delta_{\rm min}=10^{-9}$, 
$\delta_{0}=10^{-6}$,
$\delta_{\rm max}=0.1$, 
$r = 2$ and the average is over $10^4$ realizations 
for each point in the FSLE curve.
The horizontal lines indicate the values of $\lambda_s$ and
$\lambda_f$.
}
\label{fig:1}
\end{figure}

\begin{figure}[hbt]
\caption{
The FSLE, $\lambda(\delta)$, as a function of $\delta$ for the
coupled maps (\protect\ref{eq:ThreeMap}, \protect\ref{eq:ThreeMap1})
with the same parameters as in figure \protect\ref{fig:1}.
The filled squares refer to the results obtained with $\delta_{\rm min}=10^{-9}$ and 
$\delta_{0}=10^{-6}$ (same curve as in
figure \protect\ref{fig:1}). 
The filled triangles show the behavior of $\lambda(\delta)$ when $\delta_{\rm min} = \delta_{0} = 10^{-5}$
and the perturbation did not start along the most expanding direction.
}
\label{fig:2}
\end{figure}

\begin{figure}[hbt]
\caption{
The FSLE, $\lambda(\delta)$, as a function of $\delta$ for the
coupled maps (\protect\ref{eq:ThreeMap}, \protect\ref{eq:ThreeMap1}). 
The parameters are as in figure
\protect\ref{fig:1}. The filled squares
are as in figure 1. The filled triangles refer to the case where
the fast variables are replaced by a sequence of 
random numbers uniformly distributed in $[0,1]$. 
}
\label{fig:3}
\end{figure}

\begin{figure}[hbt]
\caption{
The FSLE, $\lambda(\delta)$, as a function of $\delta$ for the
coupled Lorenz systems (\protect\ref{clor}) with $R=45.92$,
$\sigma=16$, $b=4$ and coupling $\epsilon=10^{-4}$ (filled squares).
The filled triangles
indicate the results obtained when the perturbed 
trajectory evolves according to the 
modified dynamics (\protect\ref{noiselor}) with $\epsilon=10^{-4}$.
The open diamonds represent the case $\epsilon=0$ in the perturbed trajectory (no fast dynamics).
}
\label{fig:4}
\end{figure}

\begin{figure}[hbt]
\caption{
The FSLE, $\lambda(\delta)$, as a function of $\delta$ for the
time series $x(n)$ obtained from the coupled maps (\protect\ref{eq:ThreeMap}, 
\protect\ref{eq:ThreeMap1}) with $\delta_{\rm min}=10^{-5}$,
$\delta_{0}=10^{-4}$ and coupling $\epsilon=0.02$ (filled squares). 
The triangles indicate the results for an embedding
dimension $M=1$ and the diamonds for $M=2$. 
The number of realizations used for
each point in the FSLE is $10^{4}$. 
The total number of point in the time series
is $10^8$ for the case $M=1$ and more than $10^{10}$ for $M=2$.
}
\label{fig:5}
\end{figure}

\begin{figure}[hbt]
\caption{
The FSLE, $\lambda(\delta)$, as a function of $\delta$ for the
time series $x_s(t)$ obtained from the coupled Lorenz models 
(\protect\ref{clor}) with coupling $\epsilon=0.05$ (filled squares). 
The filled triangles indicate the results for
an embedding dimension $M=3$, time delay $\tau=0.02$
and a total number of points $N=500,000$. 
}
\label{fig:6}
\end{figure}


\begin{thebibliography}{99}

\bibitem{TO}
P. G. Drazin and G. P. King, Eds., 
{\it Interpretation of Time Series from
Nonlinear Systems} {\it Physica D} {\bf 58} (1992).

\bibitem{OSY}
E. Ott, T. Sauer and J. A. Yorke, {\it Coping with Chaos} (Wiley, New
York, 1994).

\bibitem{Aba}
H. D. I. Abarbanel, R. Brown, J. J. Sidorowich and L. Sh. Tsimring, 
{\it Rev. Modern Phys.} {\bf 65} (1993) 1331.

\bibitem{Kant}
H. Kantz, in {\it Nonlinear Physics of Complex Systems},
J. Parisi, S.C. M\"uller and W. Zimmermann Eds.
(Springer, Berlin, 1996), 213.

\bibitem{ER85}
J.-P. Eckmann and D. Ruelle, {\it Rev. Modern Phys.} {\bf 57} 
(1985) 617.

\bibitem{Pack}
N. Packard, J. Crutchfield, J. D. Farmer and R. Shaw, {\it Phys. Rev.
Lett.} {\bf 45} (1981) 712.

\bibitem{Tak}
F. Takens, in {\it Dynamical Systems and Turbulence, Warwick 1980},
D. Rand and L. S. Young Eds. (Springer, Berlin, 1981), p. 366.

\bibitem{GP}
P. Grassberger and I. Procaccia, {\it Phys. Rev.
Lett.} {\bf 50} (1983) 346.

\bibitem{GPe}
P. Grassberger and I. Procaccia, {\it Phys. Rev. A} {\bf 28} (1983),
2591.

\bibitem{Wolf}
A. Wolf, J. B. Swift, H. Swinney and J. A. Vastano, {\it Physica  
D} {\bf 16} (1985) 285.

\bibitem{Smith}
L. A. Smith, {\it Phys. Lett. A} {\bf 133} (1988) 283.

\bibitem{ER92}
J.-P. Eckmann and D. Ruelle, {\it Physica D} {\bf 56} (1992) 185.

\bibitem{OP}
A. R. Osborne and A. Provenzale, {\it Physica D} {\bf 35} (1989) 357.

\bibitem{Vio}
R. Vio, S. Cristiani, O. Lessi and A. Provenzale, {\it Astrophys. J.} 
{\bf 391} (1992) 518.

\bibitem{Pro}
A. Provenzale, L. A. Smith, R. Vio and G. Murante,
{\it Physica D} {\bf 58} (1992) 31.

\bibitem{PST}
N. Platt, E. A. Spiegel and C. Tresser {\it Phys. Rev. Lett.} {\bf 70} 
(1994) 279.

\bibitem{Harden}
J. Graf von Hardenberg, F. Paparella, N. Platt,
A. Provenzale, E. A. Spiegel, C. Tresser, {\it Phys. Rev. E}{\bf 50}
(1997) 58.

\bibitem{Bof}
G. Boffetta, G. Paladin and A. Vulpiani,
{\it J. Phys. A}{\bf 29}, (1996) 2291.

\bibitem{ABCPV96}
E. Aurell, G. Boffetta, A. Crisanti, G. Paladin and A. Vulpiani,
{\it Phys. Rev. Lett.} {\bf 77} (1996) 1262;
{\it J. Phys. A}{\bf 30} (1997) 1.

\bibitem{BGPV97}
G. Boffetta, P. Giuliani, G. Paladin and A. Vulpiani, {\it J. Atmos.
Sci.} in press (1997).

\bibitem{Lorenz96}
E.N. Lorenz, 
Predictability - a problem partly solved. {\it Proc. Seminar on
predictability,} Reading, U.K., European Centre for Medium-Range Weather
Forecast, (1996) 1.

\bibitem{Kol56}
A.N. Kolmogorov, {\it IRE Trans. Inf. Theory} {\bf 1} (1956) 102.

\bibitem{GW93}
P. Gaspard and X.J. Wang, {\it Physics Reports} {\bf 235} (1993) 291.

\bibitem{Lorenz63}
E. Lorenz, {\it J. Atmos. Sci.} {\bf 20} (1963) 130.

\bibitem{CR}
J. Graf von Hardenberg, F. Paparella, A. Provenzale, E. A. Spiegel, in
{\it Nonlinear Signal and Image Analysis}, J. R. Buchler and
H. Kandrup Eds. (Annals N.Y. Acad. Sci., 1997), p. 79.

\end{thebibliography}
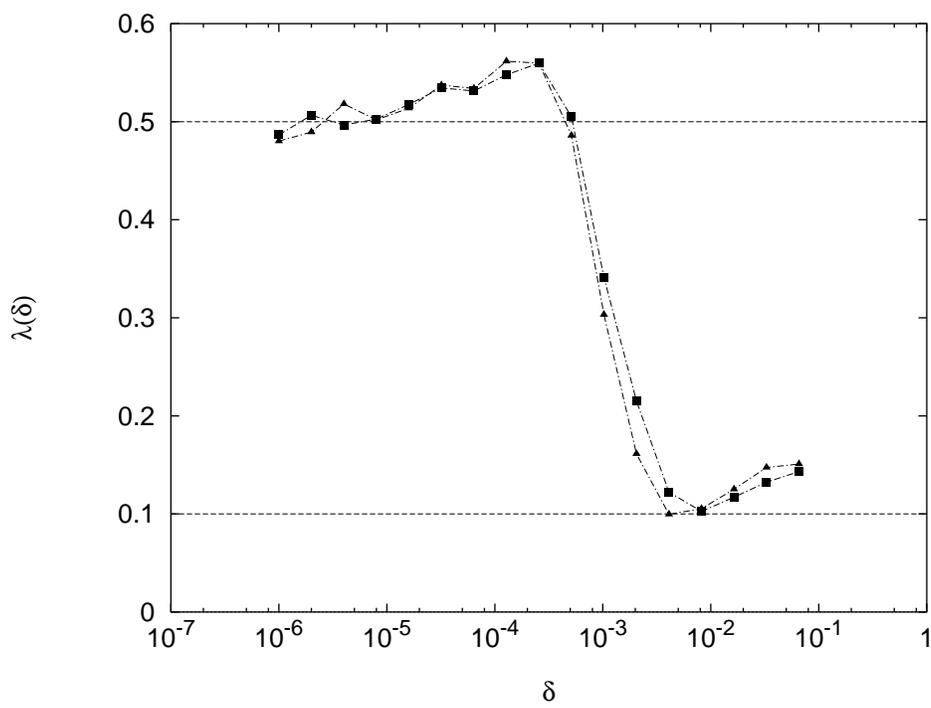
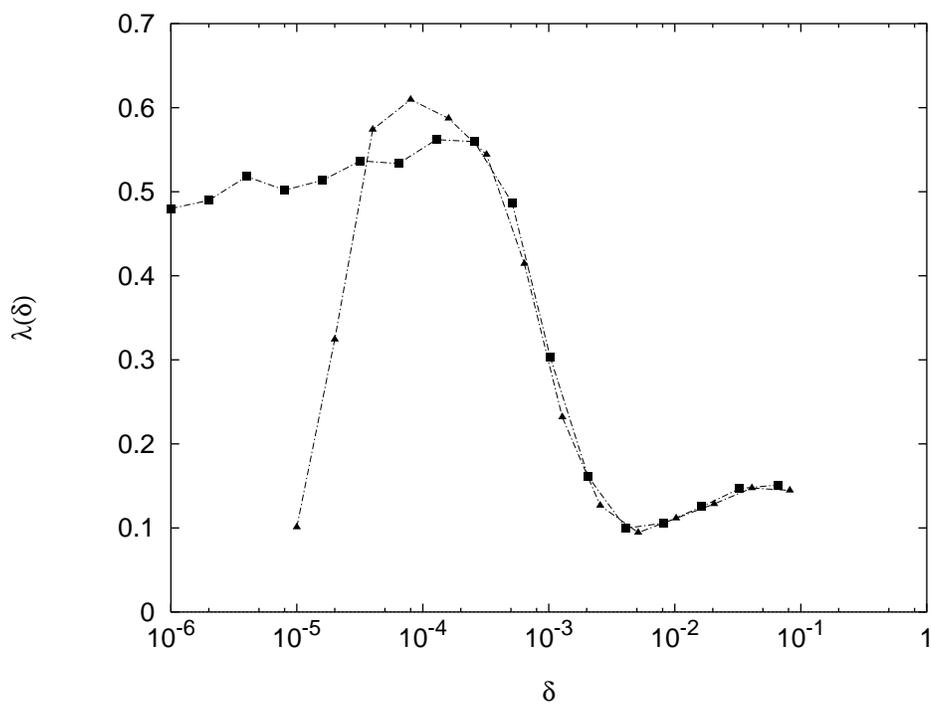
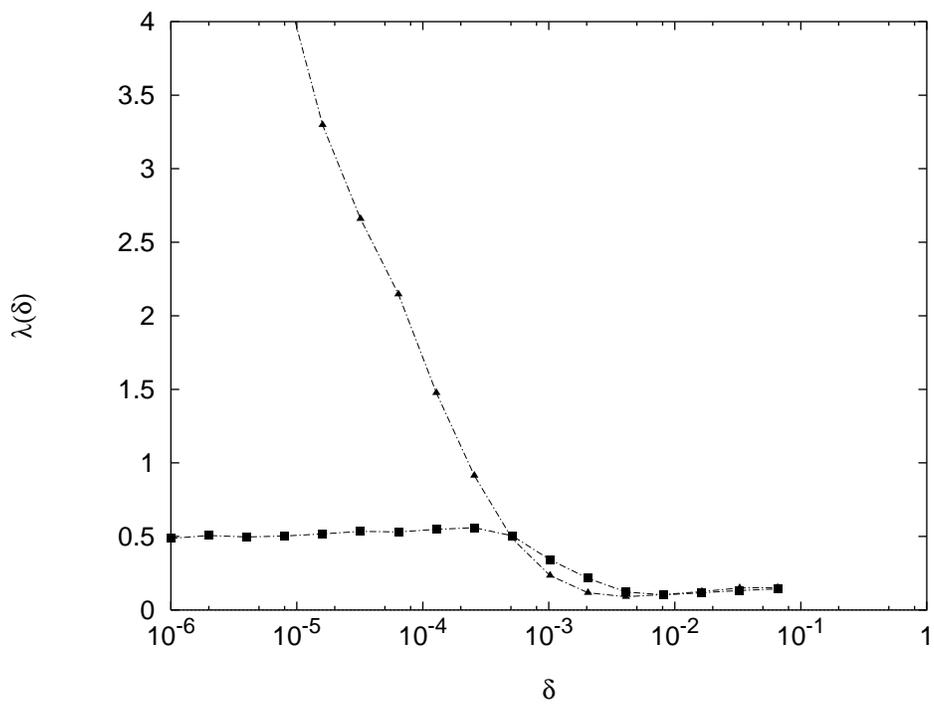
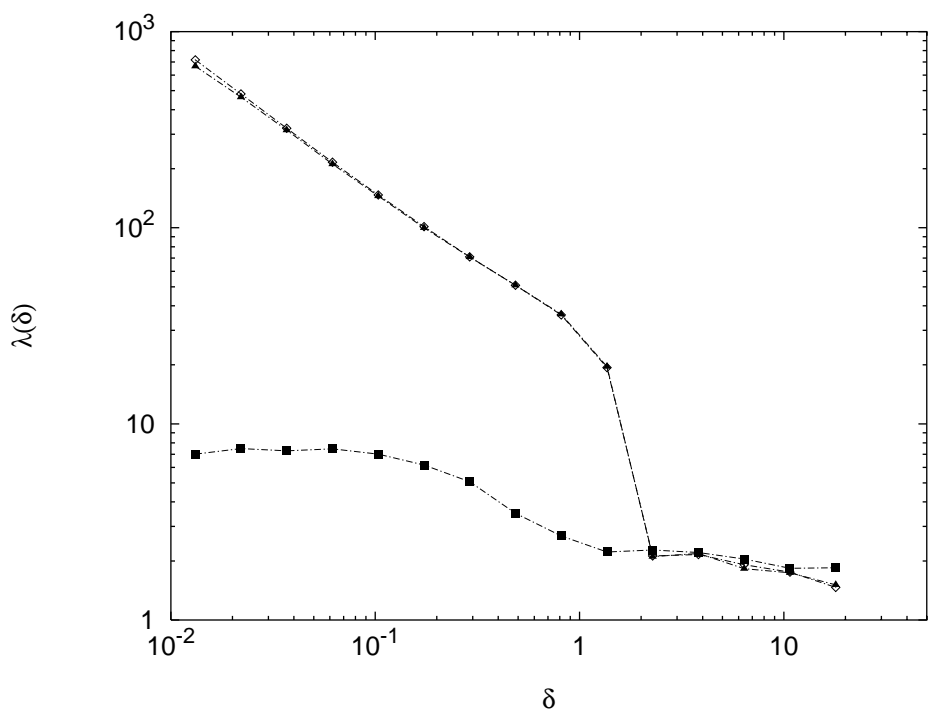
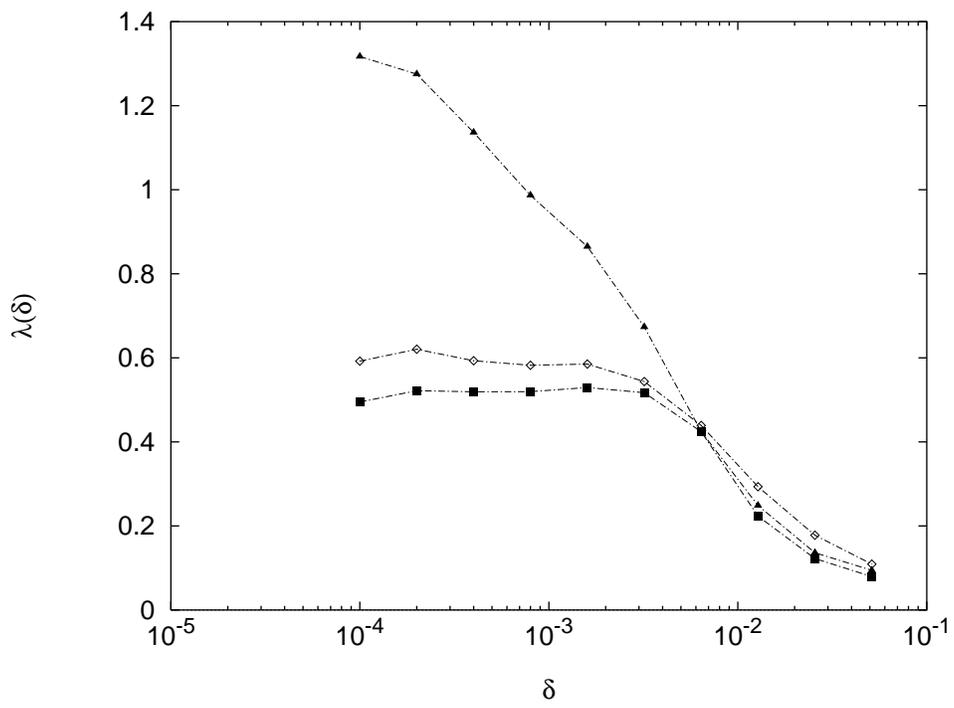
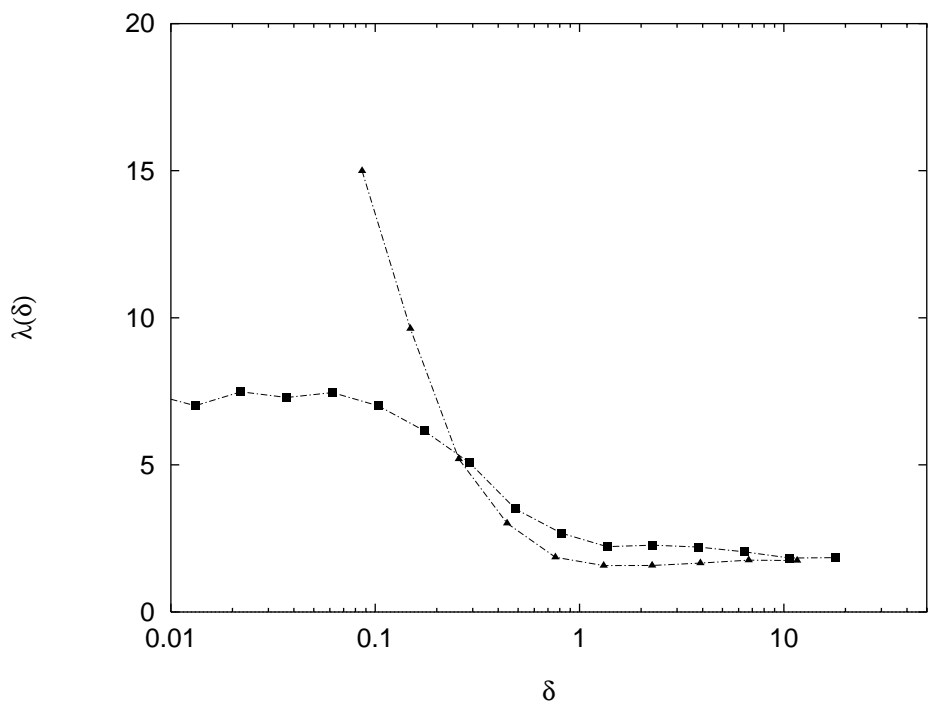
\end{document}